\documentclass[12pt,a4paper]{article}
\pdfoutput=1
\usepackage{amsthm,bm,amsmath,amssymb,epsfig,multicol,multirow}
\usepackage[round]{natbib}
\usepackage{graphics,rotating,graphicx}
\usepackage{array,multirow,caption2}
\usepackage{amssymb}
\usepackage{amsthm}
\usepackage{bm,amsmath,multicol}

\newcommand{\argmin}[1]{\underset{#1}{\operatorname{argmin}}}

\newtheorem{theorem}{Theorem}

\newtheorem{condition}{Condition}
\newtheorem{model}{Model}
\newtheorem{method}{Method}

\begin{document}

\noindent{\Large \textbf{Simple response and predictor transformations to adjust for symmetric dependency in dimension reduction for visualization}}



\vspace{0.5cm}

\noindent LUKE A. PRENDERGAST

\noindent \textit{Department of Mathematics and Statistics, La Trobe
University}

\vspace{0.5cm}

\noindent ALEXANDRA L. GARNHAM

\noindent \textit{Department of Mathematics and Statistics, La Trobe
University}
\vspace{1cm}

\noindent\textbf{ABSTRACT.  In the regression setting, dimension reduction allows for complicated regression structures to be detected via visualization in a low-dimension framework.  However, some popular dimension reduction methodologies fail to achieve this aim when faced with a problem often referred to as symmetric dependency.  In this paper we show how vastly superior results can be achieved when carrying out response and predictor transformations for methods such as least squares and Sliced Inverse Regression.  These transformations are simple to implement and utilize estimates from other dimension reduction methods that are not faced with the symmetric dependency problem.  We highlight the effectiveness of our approach via simulation and an example.  Furthermore, we show that ordinary least squares can effectively detect multiple dimension reduction directions.  Methods robust to extreme response values are also considered.}

\vspace{0.5cm}



\section{Introduction}

Let $Y \in \mathbb{R}$ denote a random univariate response and $\mathbf{x}\in\mathbb{R}^p$ a random $p$-dimensional vector of predictors.  \cite{LI&DU89} considered the model
\begin{equation}
Y = f(\bm{\beta}^\top\mathbf{x},\varepsilon)\label{model}
\end{equation}
where $\bm{\beta}$ is an unknown $p$-dimensional vector of predictor coefficients, $f$ is the unknown link function and $\varepsilon$ is the error term that is assumed to be independent of $\mathbf{x}$.  Of interest is the regression function $E(Y|\mathbf{x})$ where, ideally, a plot of $Y$ versus $\mathbf{x}$ can reveal the form of $f$.  However, we are limited in this sense when $p$ is large due to our inability to visualize objects in high dimensions.  Importantly, $Y$ depends on $\mathbf{x}$ only through $\bm{\beta}^\top\mathbf{x}$ so that if we could determine $\bm{\beta}$ then we could replace the $p-$dimensional $\mathbf{x}$ with the one-dimensional $\bm{\beta}^\top\mathbf{x}$.  Our ability to explore possibilities for $f$ would then be enhanced due to the resulting lower-dimensional framework.

In the sample setting, let $\{y_i,\mathbf{x}_i\}^n_{i=1}$ be $n$ sample realizations of $Y$ and $\mathbf{x}$ where the relationship between $Y$ and $\mathbf{x}$ is assumed to be of the form given in \eqref{model}.  Suppose that $\bm{\beta}$ can be estimated and let this estimate be denoted $\widehat{\bm{\beta}}$.  Then the $y_i$'s can be plotted against the $\widehat{\bm{\beta}}^\top\mathbf{x}_i$'s to visually determine $f$.  Such a plot is called an \textit{Estimated Sufficient Summary Plot} \cite[ESSP, see, for e.g.,][]{CO98}.  The focus of our work here will be to obtain good ESSP's in settings for which estimation of $\bm{\beta}$ is difficult.

\cite{LI&DU89} extended earlier works by \cite{BR77, BR83} to show that ordinary least squares (OLS), and robust versions, can be used to estimate the direction of $\bm{\beta}$ when the model is of the form as in \eqref{model} and under some mild conditions for $\mathbf{x}$.  We will provide a brief review of these results in Section 2.  However, for some forms of $f$ least squares is not expected to find $\bm{\beta}$.  Consequently we also discuss another approach, Principal Hessian Directions \citep[PHD,]{LI92}, which is not restricted by these particular models.  In Section 3 we propose a simple transformation of the response based on an initial PHD estimate that can be used to ensure that OLS can provide a good ESSP.  Simulations are provided in Section 4 which highlight that this approach can be used to obtain vastly superior estimates.  Extensions are discussed in Section 5 to consider other approaches.  Finally, an example is provided in Section 6 and the paper is concluded with a discussion in Section 7.

\section{Methods}

Consider the following condition commonly referred to as the \textit{Linear Design Condition} considered by \cite{LI&DU89}.
\begin{condition}\label{LDC}
For any $\mathbf{c} \in \mathbb{R}^p$, $E(\mathbf{c}^\top\mathbf{x}|\bm{\beta}^\top\mathbf{x})=c_0+c_1\bm{\beta}^\top\mathbf{x}$ for some scalar constants $c_0$ and $c_1$.
\end{condition}
\cite{LI&DU89} highlight that this condition is satisfied when the distribution of $\mathbf{x}$ belongs to the family of elliptically symmetric distributions.  However, there are other situations for which this holds and \cite{HA&LI93} show that Condition \ref{LDC} often approximately holds in practice when $p$ is large.  One also has the possibility to utilize predictor transformations to ensure that it approximately holds \cite[see, for e.g.,][]{FO&WE11}.

\subsection{Least squares and similar approaches}

When Condition \ref{LDC} and the model in \eqref{model} hold, \cite{LI&DU89} show that the OLS slope vector, which is denoted $\mathbf{b}=\left[\text{Var}(\mathbf{x})\right]^{-1}\text{Cov}(\mathbf{x},Y)$, satisfies $c\bm{\beta}$ for a $c\in \mathbb{R}$.  Consequently, OLS can recover the direction of $\bm{\beta}$ when $c\neq 0$ and a plot of $Y$ versus $\mathbf{b}^\top\mathbf{x}$ used to seek $f$.  It should be pointed out that any nonzero $c$ is adequate since any $\mathbf{b}$ in the direction of $\bm{\beta}$ is suitable for finding an appropriate link function.  In practice , OLS can be used to obtain $\widehat{\mathbf{b}}$, the usual OLS slope estimate, and an ESSP created using the $y_i$'s and the $\widehat{\mathbf{b}}^\top\mathbf{x}_i$'s.  While OLS is one simple approach, Li and Duan's results are generalized to include estimators satisfying
\begin{equation}
\argmin{a,\mathbf{b}}\frac{1}{n}\sum^n_{i=1}\rho(a+\mathbf{b}^\top\mathbf{x}_i,y_i)\label{rho}
\end{equation}
provided that $\rho$ is convex in its first argument and that a solution exists.  Hence, other possibilities can be robust estimators such as $M$-estimators with the Huber weight function \citep{HU73}.  While the temptation would be to only consider a robust approach when possible errors are present in the data set, \cite{PR&SH13} showed that for some models the robust estimators can outperform OLS even when data is sampled without error.  Similarly, \cite{PR08} used to trimming of influential observations to also improve estimates.

If an estimator is expected to find the direction of $\bm{\beta}$, then it is required that $c\neq 0$.  Some discussion of when this does not occur (i.e. when $c=0$) can be found in \cite{LI91} and \cite{CO&WE91}.  While these discussions are for a different method they can similarly be applied to OLS.  That is, when the link function $f$ is symmetric about the mean of $\bm{\beta}^\top\mathbf{x}$, then $\mathbf{b}=\mathbf{0}$.  To highlight this, we consider two simulated examples; the first does not have the symmetric dependency issue while the second does. The models we will use are
\begin{model} $Y = \sin(0.5\bm{\beta}^\top\mathbf{x})+0.05\varepsilon$\label{sine_model}\end{model}
\begin{model} $Y = \cos(0.5\bm{\beta}^\top\mathbf{x})+0.05\varepsilon$\label{cos_model}\end{model}
\noindent where, for both models, $\mathbf{x}\sim N_{10}(\mathbf{0},\mathbf{I}_{10})$, $\varepsilon\sim N(0,1)$ and $\bm{\beta}=[1,-2,0,\ldots,0]^\top$.

\begin{figure}[h!t]
\centering
\includegraphics[scale = 0.9]{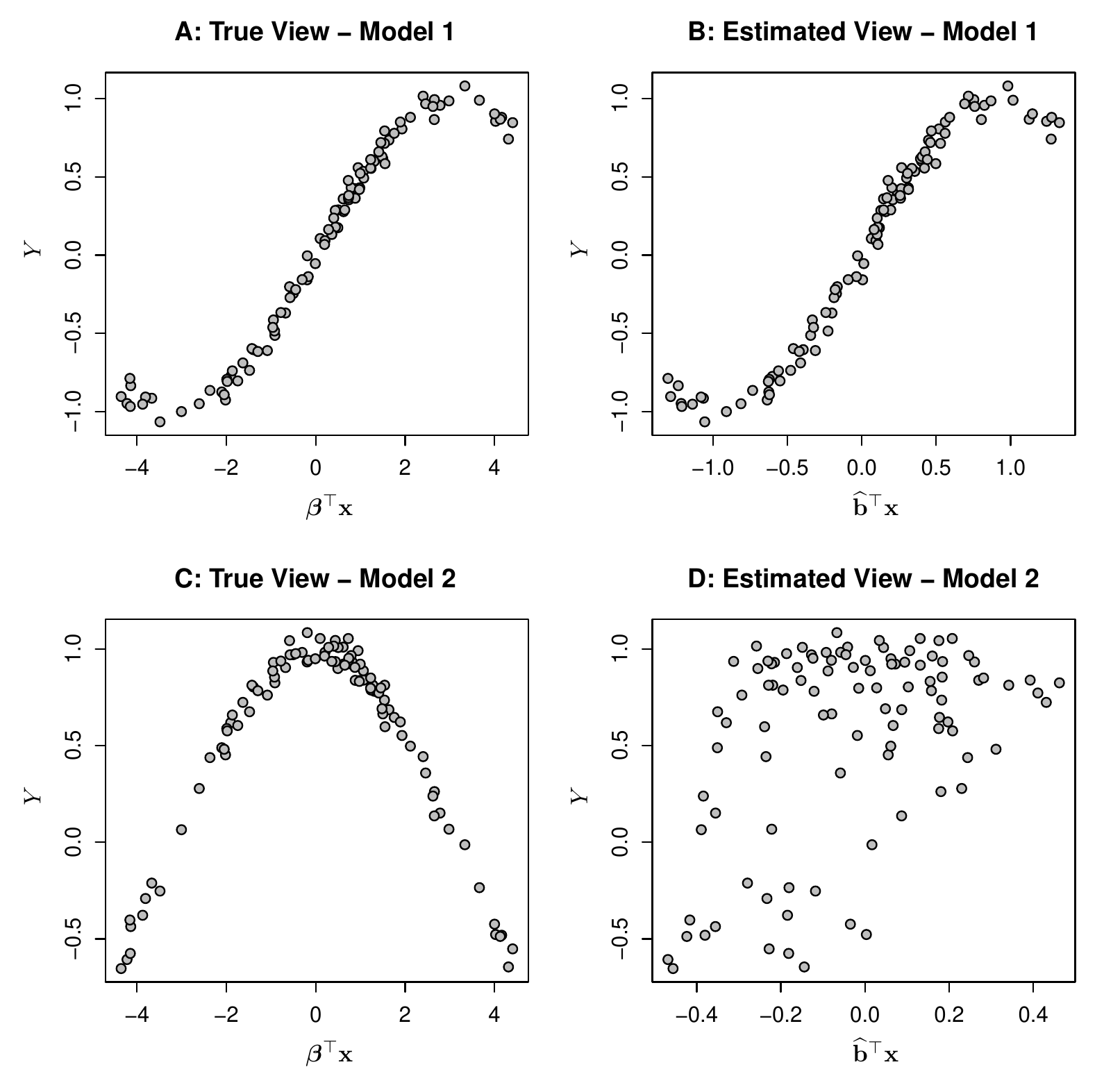}\caption{Plots of $y_i$'s versus the $\bm{\beta}^\top\mathbf{x}_i$'s (True Views) and $y_i$'s versus the $\widehat{\mathbf{b}}^\top\mathbf{x}_i$'s (Estimated Views - ESSPs) for 100 observations generated for Model 1 (Plots A and B) and Model 2 (Plots C and D).  OLS was used to estimate the direction of $\bm{\beta}$.}\label{figure:1}
\end{figure}

In Figure \ref{figure:1} we provide true views (where the $y_i$'s are plotted against the ideally dimension reduced $\mathbf{x}_i$'s - i.e. the $\bm{\beta}^\top\mathbf{x}_i$'s) and ESSP's where OLS has been used as the estimator.  Plots A and B are for Model 1 and Plots C and D for Model 2 where, in both cases, $n=100$ observations have been randomly generated.  If OLS has performed well, then we would expect the ESSP is look similar to the true views with possible differences in scale on the horizontal axis since OLS is targeting $c\bm{\beta}$ for a $c$ that is not necessarily one.  For Model 1, we can see that OLS has performed exceptionally well producing an excellent ESSP as seen in Plot B.  However, OLS has failed for Model 2 with an ESSP in Plot D that does not provide any evidence of a relationship between the responses and dimension reduced predictors.  The true view though shows that there is certainly something to find.  Recall that Model 2 exhibits symmetric dependency and OLS is trying to estimate $0\times \bm{\beta}$.

\subsection{Principal Hessian Directions}

\cite{LI92} introduced Principal Hessian Directions (PHD) - a method that does not suffer from the symmetric dependency problem and one that is also capable of finding multiple vectors of predictor coefficients.  That is, the model can be assumed to be of the form
\begin{equation}
Y = f(\bm{\beta}_1^\top\mathbf{x},\ldots, \bm{\beta}_K^\top\mathbf{x},\varepsilon)\label{Kmodel}
\end{equation}
in which case it is desirable to find a basis for the span of the $\bm{\beta}_k$'s.

Consider the following condition required by PHD.
\begin{condition}\label{normal}
$\mathbf{x}\sim N_p(\bm{\mu}, \bm{\Sigma})$.
\end{condition}
When Condition \ref{normal} holds imposing normality of the predictor, Condition \ref{LDC} also holds.  As a consequence, if PHD is applicable due to this condition being met then so to is OLS.  There are slightly weaker conditions for PHD to work, namely that $\text{Var}(\mathbf{x}|\bm{\beta}^\top\mathbf{x})$ is constant in conjunction with assuming Condition \ref{LDC} holds. Recent work by \cite{LE13} show that this will often hold approximately in practice.

While OLS returns an estimate of $\bm{\Sigma}^{-1}\bm{\Sigma}_{xy}$ (where $\bm{\Sigma}_{xy}$ is the covariance vector between $\mathbf{x}$ and $Y$) as an estimate of the direction of $\bm{\beta}$, PHD instead carries out an eigen-decomposition of an estimate to $\overline{\mathbf{H}}=\bm{\Sigma}^{-1}\bm{\Sigma}_{yxx}\bm{\Sigma}^{-1}$ where
$$\bm{\Sigma}_{yxx}=E\left[\{Y-E(Y)\}(\mathbf{x}-\bm{\mu})(\mathbf{x}-\bm{\mu})^\top\right].$$
For \textit{many models} satisfying \eqref{model} and when Condition \ref{normal} holds, the rank of $\overline{\mathbf{H}}$ is one and the eigenvector corresponding to the non-zero eigen-value is in the same direction as $\bm{\beta}$.  We have emphasized many models here since PHD will not be able to find the direction of $\bm{\beta}$ when there is odd symmetric dependency between $Y$ and the mean of $\bm{\beta}^\top\mathbf{x}$.   Here odd symmetric dependency refers to the type of symmetry seen for Model 1 (see Figure \ref{figure:1}).  \cite{LI91} also noted that $Y$ can be replaced by the OLS residual without changing $\overline{\mathbf{H}}$ where, notationally, we replace $\bm{\Sigma}_{yxx}$ by $\bm{\Sigma}_{rxx}$ to distinguish between the two approaches.  While $\overline{\mathbf{H}}$ does not change, the estimator is influenced.  Empirical and theoretical results have suggested that this residual based PHD approach is often a better estimator of $\bm{\beta}$ \citep{CO98a, PR&SM10}.  Consequently the residuals-based PHD will be our method of choice.

Since $\bm{\Sigma}_{rxx}$ is moment-based, estimation is straightforward.  Let $r_1,\ldots,r_n$ denote the usual OLS residuals for the regression of the $y_i$'s on the $\mathbf{x}_i$'s and also let $\overline{\mathbf{x}}$ be the sample mean of the $\mathbf{x}_i$'s.  Then the estimate to $\bm{\Sigma}_{rxx}$ is
$$\widehat{\bm{\Sigma}}_{rxx}=\frac{1}{n}\sum^n_{i=1}r_i(\mathbf{x}_i-\overline{\mathbf{x}})(\mathbf{x}_i-\overline{\mathbf{x}})^\top.$$

Similarly to OLS estimation, \cite{LU01} has previously shown that trimming can improve PHD estimation.

\section{Predictor and response transformations to remove symmetric dependency}

\cite{GA&PR13,GA&PR13a} show that response transformations can greatly improve OLS and PHD estimates.  Their results, however, do not solve the issue of the symmetric dependency that troubles OLS.  The aim here is to introduce two transformation functions, one for the response and the other for the predictor, that can be useful in the symmetric dependency setting.

\subsection{Theory}\label{subsection:theory}

The response transformation that we will be focusing on is
\begin{equation}
t_y(Y; \mathbf{v}) = \left\{ \begin{array}{ll} Y,& \mathbf{v}^\top\mathbf{x} > \mathbf{v}^\top\bm{\mu} \\ Y - 2[Y - E(Y|\mathbf{v}^\top\mathbf{x}=\mathbf{v}^\top\bm{\mu})],& \mathbf{v}^\top\mathbf{x} \leq \mathbf{v}^\top\bm{\mu}\end{array}\right.\label{ty}
\end{equation}
where $\mathbf{v}$ needs to be chosen.  If $\mathbf{v}=c_1\bm{\beta}$ for a nonzero scalar $c_1$, then $t_y(Y; \mathbf{v})$ and $\mathbf{x}$ still satisfy the model in \eqref{model} and Condition \ref{LDC}.  An estimator of $\bm{\beta}$ is then the OLS slope vector estimator for the regression of $t_y(Y; \mathbf{v})$ on $\mathbf{x}$.  Soon we will show that in the empirical setting, good estimates to $\bm{\beta}$ used in the transformation function can generate improved results.  We also consider the following predictor transformation function
\begin{equation}
t_x(\mathbf{x}-\bm{\mu}; \mathbf{v}) = \text{sign}(\mathbf{v}^\top\mathbf{x} - \mathbf{v}^\top\bm{\mu})(\mathbf{x}-\bm{\mu}).\label{tx}
\end{equation}

\begin{figure}[h!t]
\centering
\includegraphics[scale = 0.9]{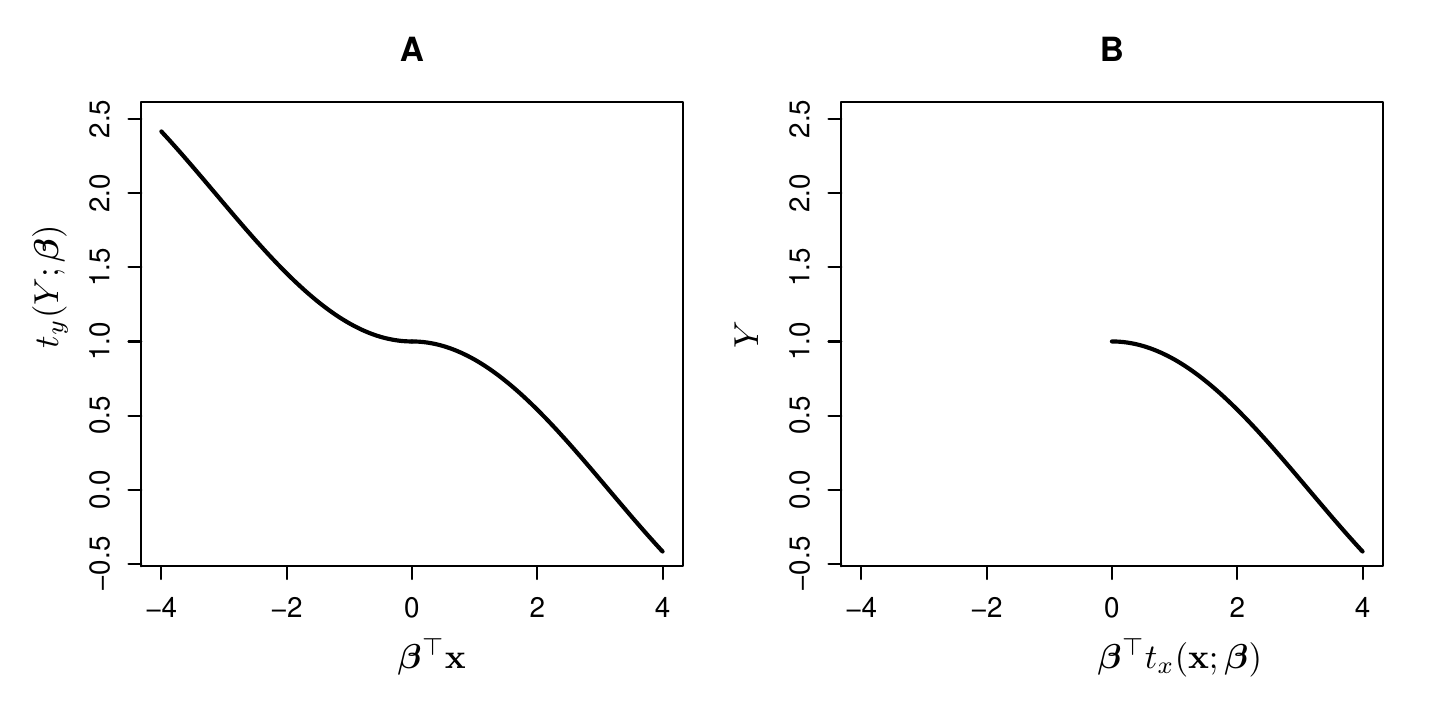}\caption{Under the assumption of zero error in the Model 2 and choosing $\mathbf{v}=\bm{\beta}$, Plot A provides the plot of the transformed $Y$ versus $\bm{\beta}^\top\mathbf{x}$ and Plot B provides the plot of $Y$ versus the transformed  $\bm{\beta}^\top t_x(\mathbf{x}-\bm{\mu}; \mathbf{v})$.}\label{figure:2}
\end{figure}
In Figure 2 we show the effects of the transformations on the curves defining $Y$ in Model 2 under the assumption of zero error.  In Plot A it is clear now that the transformed $Y$ is no longer symmetric about the mean of $\bm{\beta}^\top\mathbf{x}$ (zero).  In Plot B it is clear that $Y$ is not symmetric about the mean of $\bm{\beta}^\top t_x(\mathbf{x}-\bm{\mu}; \mathbf{v})$, alleviating the symmetric dependency problem.  For this latter plot, the transformation folds the curve back on to itself (left-to-right).

In the theorem below, we identify that for certain choices of $\mathbf{v}$, the transformation in \eqref{tx} can be used to find the direction of $\bm{\beta}$.  The proof is in the Appendix.
\begin{theorem}\label{theorem:tx(v)proof}
Consider the predictor transformation considered in \eqref{tx} and let $\mathbf{v}=c_1\bm{\beta}$ for any $c_1\in \mathbb{R}$.  Under the model in \eqref{model} and Condition \ref{LDC},
\begin{align}
\left[\text{Var}(\mathbf{x})\right]^{-1}\text{Cov}\left[t_x(\mathbf{x}-\bm{\mu}; \mathbf{v}), Y\right] = &c_2\bm{\beta}\label{ols-1}\\
\left\{\text{Var}[t_x(\mathbf{x}-\bm{\mu}; \mathbf{v})]\right\}^{-1}\text{Cov}\left[t_x(\mathbf{x}-\bm{\mu}; \mathbf{v}), Y\right] = & c_3\bm{\beta}\label{ols-2}
\end{align}
for constant scalars $c_2,c_3\in\mathbb{R}$.
\end{theorem}

The second estimator provided in \eqref{ols-2} is simply the OLS slope from the regression of $Y$ on $t_x(\mathbf{x}-\bm{\mu}; \mathbf{v})$.  The first, provided in \eqref{ols-1}, is similar although utilizes the variance estimator for the original $\mathbf{x}$.

\subsection{Application in practice}\label{section:methods}

Recall that our sample of $n$ observations are denoted $\{y_i,\mathbf{x}_i\}^n_{i=1}$.  Throughout let $\overline{y}$, $\overline{\mathbf{x}}$, $\mathbf{S}_{x}$ and $\mathbf{S}_{xy}$ denote the sample mean of the $y_i$'s, sample mean of the $\mathbf{x}_i$'s, sample covariance matrix of the $\mathbf{x}_i$'s and the sample covariance between the $\mathbf{x}_i$'s and $y_i$'s respectively.  Also let $\mathbf{X}$ denote the $n\times p$ design matrix whose $i$th row is $\mathbf{x}_i$.

There are two points that need clarification prior to application in practice.  Firstly, how to choose an appropriate vector $\mathbf{v}$?  Our simulations indicate that for many models, OLS is a better estimator of $\bm{\beta}$ in \eqref{model} than PHD.  However, PHD is preferred when symmetric dependency is evident in which case OLS can struggle to find $\bm{\beta}$.  Consequently, in practice we propose to set $\mathbf{v}=\widehat{\mathbf{b}}_{phd}$ - the PHD estimate to $\bm{\beta}$.  In the next section our results show that reasonable PHD estimates of $\bm{\beta}$ can lead to much-improved estimates to $\bm{\beta}$ when OLS is employed following the transformations in Section \ref{subsection:theory}.

Secondly, the response transformation requires estimation of $E(Y|\mathbf{v}^\top\mathbf{x}=\mathbf{v}^\top\bm{\mu})$.  We propose to find an approximation to this estimate as
\begin{equation}
\overline{y}({\mathbf{v}}) = \frac{1}{m}\sum_{j \in I_m}y_j
\end{equation}
where $I_m$ is the set of indices for the closest $m$ $\mathbf{v}^\top\mathbf{x}_i$'s to $\mathbf{v}^\top\overline{\mathbf{x}}$.  In what follows we arbitrarily set $m=10$ and obtain good results.

The transformations we will employ are then
\begin{equation}
y_i^*=t_{y_i}(y_i; \widehat{\mathbf{b}}_{phd}) = \left\{ \begin{array}{ll} y_i,& \widehat{\mathbf{b}}_{phd}^\top\mathbf{x}_i > \widehat{\mathbf{b}}_{phd}^\top\overline{\mathbf{x}} \\ y_i - 2[y_i - \overline{y}(\widehat{\mathbf{b}}_{phd})],& \widehat{\mathbf{b}}_{phd}^\top\mathbf{x}_i \leq \widehat{\mathbf{b}}_{phd}^\top\overline{\mathbf{x}}\end{array}\right.\label{sample-ty}
\end{equation}
and
\begin{equation}
\mathbf{x}_i^*=t_{x_i}(\mathbf{x}_i-\overline{\mathbf{x}}; \widehat{\mathbf{b}}_{phd}) = \text{sign}(\widehat{\mathbf{b}}_{phd}^\top\mathbf{x}_i - \widehat{\mathbf{b}}_{phd}^\top\overline{\mathbf{x}})(\mathbf{x}_i-\overline{\mathbf{x}})\label{sample-tx}
\end{equation}
where we will use the notations $y_i^*$ and $\mathbf{x}_i^*$ for convenience.  The methods we will use are:
\begin{method}\label{m1}
The OLS slope vector for the regression of the $y_i^*$'s on the $\mathbf{x}_i$'s.
\end{method}
\begin{method}\label{m2}
The OLS slope vector for the regression of the $y_i$'s on the $\mathbf{x}_i^*$'s.
\end{method}

Potentially, a combination of the transformations could also be used.  However, our simulations revealed the better results are achieved by using only one at a time.  Additionally, another possibility exists and that is to use $\mathbf{S}^{-1}_x\mathbf{S}^*_{xy}$ .  However, our simulations also revealed that this approach was very typically inferior to the other two.  For brevity, we therefore do not consider this approach further.

\subsection{An iterative approach}

A potential problem with the transformation methods is that the initial estimated direction is poor.  However, one approach is to alleviate this is to apply an iterative scheme which starts with the initial estimate, obtains a new estimate after transformation and iteratively uses the new estimate as the initial estimate until convergence.  Hence, a general algorithm for this approach is:
\begin{description}
\item[Step 0.1:] Estimate the direction of $\bm{\beta}$ using PHD and denote this as $\widehat{\mathbf{b}}^{(1)}$.
\item[Step 0.2:] Set $i=1$ and tol.met $=$ FALSE.
\item[Step $\bm{i}$:] While tol.met is FALSE do
\begin{description}
\item[Step $\bm{i}$.1:] Apply \textbf{Method j} using $\widehat{\mathbf{b}}^{(i)}$ as the direction for transformation and obtain a new estimated direction $\widehat{\mathbf{b}}^{(i+1)}$.
\item[Step $\bm{i}$.2:] If $1-\text{cor}^2(\mathbf{X}\widehat{\mathbf{b}}^{(i)}, \mathbf{X}\widehat{\mathbf{b}}^{(i+1)})<\text{tol}$ then set tol.met to TRUE.
\item[Step $\bm{i}$.3:] Increment $i = i + 1$.
\end{description}
\item[Step $\bm{i+1}$:] Return $\widehat{\mathbf{b}}^{(i)}$ as the final estimate to the direction of $\bm{\beta}$.
\end{description}

We have chosen $\text{cor}^2(\mathbf{X}\widehat{\mathbf{b}}^{(i)}, \mathbf{X}\widehat{\mathbf{b}}^{(i+1)})$ as the criterion for exiting the iterating loop since, when there is correlation present amongst the columns of $\mathbf{X}$, notably different directions can result in very similar ESSP's - which is the targeted estimate.  However, substantial differences in the squared correlation will be similarly be noted by changes in the ESSP.  Such an assessment is often used; for example, \cite{LI91} uses the squared trace correlation which is a multi-index version to compare collections of estimated directions in dimension reduction.

\section{Simulations}

In this section we consider the performance of the transformation approaches referred to as Methods 1 and 2 defined earlier.  Comparisons are also made with standard OLS and PHD estimation before we consider other methods in the next section.

\begin{table}[h!t]
\centering
\begin{tabular}{lllllllll}
  \hline
  Method & $p=10$ &  &  &    & $p=20$ &  &  &    \\
   & $n=50$ & $n=100$ & $n=200$ & $n=500$ & $n=50$ & $n=100$ & $n=200$ & $n=500$ \\  \hline
  OLS &          0.220  &         0.213  &        0.211  &        0.210    &         0.130 &         0.123 &         0.122 &         0.122    \\
      &  \textit{0.219} & \textit{0.211} &\textit{0.208} &\textit{0.207}   & \textit{0.150}& \textit{0.140}& \textit{0.137}& \textit{0.137}   \\
  PHD &          0.815  &         0.921  &        0.964  &        0.987    &         0.456 &         0.792 &         0.915 &         0.970    \\
      &  \textit{0.103} & \textit{0.041} &\textit{0.018} &\textit{0.006}   & \textit{0.197}& \textit{0.082}& \textit{0.030}& \textit{0.010}   \\
  M1  &          0.948  &         0.990  &        0.997  &        0.999    &         0.609 &         0.950 &         0.991 &         0.997    \\
      &  \textit{0.078} & \textit{0.008} &\textit{0.002} &\textit{0.001}   & \textit{0.273}& \textit{0.050}& \textit{0.005}& \textit{0.001}   \\
  M1-it  &          0.975  &         0.994  &        0.997  &        0.999    &         0.683 &         0.985 &         0.994 &         0.998    \\
      &  \textit{0.062} & \textit{0.003} &\textit{0.001} &\textit{0.001}   & \textit{0.293}& \textit{0.028}& \textit{0.002}& \textit{0.001}   \\
  M2  &          0.948  &         0.989  &        0.996  &        0.999    &         0.613 &         0.948 &         0.989 &         0.997    \\
      &  \textit{0.076} & \textit{0.010} &\textit{0.002} &\textit{0.001}   & \textit{0.271}& \textit{0.052}& \textit{0.006}& \textit{0.001}   \\
  M2-it  &          0.976  &         0.997  &        0.999  &        1.000    &         0.643 &         0.988 &         0.997 &         0.999   \\
      &  \textit{0.069} & \textit{0.002} &\textit{0.001} &\textit{0.000}   & \textit{0.283}& \textit{0.034}& \textit{0.001}& \textit{0.000}   \\
  \hline
\end{tabular}
\caption{Average $\text{cor}^2(\mathbf{X}\bm{\beta},\mathbf{X}\widehat{\mathbf{b}})$ across 10,000 simulated runs for Model \ref{cos_model} with different choices of $n$ and $p$ and where
 $\widehat{\mathbf{b}}$ is the estimate from one of five methods; OLS, PHD, Method 1 (M1) and Method 2 (M2).  M1-it and M2-t refer to the iterative estimation scheme for Methods 1 and 2.  Standard deviations are in italics.}\label{table:sim1}
\end{table}

In Table \ref{table:sim1} we provide simulated average squared correlations (with standard deviations in italics) for Model \ref{cos_model} over 10,000 runs between $\mathbf{X}\bm{\beta}$ and $\mathbf{X}\widehat{\mathbf{b}}$ - the true and estimated dimension reduced predictors.  The estimators considered are OLS, PHD and the transformation approaches. Both $p=10$ and $p=20$ were considered.  As expected OLS performs poorly due the symmetric dependency evident in the model while PHD performs well.  However, Methods 1 and 2 perform exceptionally well having successfully drawn on the good PHD estimates to remove the symmetric dependency.  The iterative estimation methods also provide improved estimates.  We exited the iterative procedure when $\text{cor}^2(\mathbf{X}\widehat{\mathbf{b}}^{(i)}, \mathbf{X}\widehat{\mathbf{b}}^{(i+1)})\geq 0.999$ or when ten iterations were reached.  The small standard deviations for the methods indicate consistently excellent results.

\cite{PR&SH13} found that robust least squares regression methods can provide improved single-index model estimates even when the data is well-behaved in the sense it has been sampled from a single index model and with a normal $\mathbf{x}$.  Similarly, \cite{PR08} found that trimming observations in the estimation step can also improve outcomes.  We further explore this by considering the following model:
\begin{model} $Y = 1/|\bm{\beta}^\top\mathbf{x}|+0.05\varepsilon$ with $p=20$ and $\bm{\beta}=[1,-2,0,\ldots,0]^\top$.\label{large_y_model}\end{model}
For this model we will assume that $\mathbf{x}\sim N(\mathbf{0},\mathbf{I}_{20})$ so that this model also consists of the symmetric dependency that troubles OLS.  Also, data simulated form the model can result in exceptionally extreme responses since the denominator in the first term on the right hand side can be very close to zero.

\begin{table}[h!t]
\centering
\begin{tabular}{llllllll}
  \hline
   Method & $n=100$ & $n=200$ & $n=500$  & Method & $n=100$ & $n=200$ & $n=500$  \\  \hline
  OLS &          0.006  &         0.004  &        0.002  & M3       &         0.202  &         0.229  &        0.259    \\
      &  \textit{0.012} & \textit{0.009} &\textit{0.005} &          & \textit{0.150} & \textit{0.147} &\textit{0.150}   \\
  RR&          0.034  &         0.035  &        0.034  & RM1      &         0.867  &         0.970  &        0.989    \\
      &  \textit{0.049} & \textit{0.049} &\textit{0.048} &          & \textit{0.130} & \textit{0.015} &\textit{0.004}   \\
  PHD &          0.801  &         0.947  &        0.985  & RM2      &         0.875  &         0.961  &        0.981    \\
      &  \textit{0.131} & \textit{0.022} &\textit{0.005} &          & \textit{0.126} & \textit{0.017} &\textit{0.006}   \\
  M1  &          0.448  &         0.638  &        0.762  & M1-trim  &         0.771  &         0.933  &        0.970    \\
      &  \textit{0.219} & \textit{0.328} &\textit{0.294} &          & \textit{0.201} & \textit{0.082} &\textit{0.031}   \\
  M2  &          0.448  &         0.594  &        0.661  & M2-trim  &         0.722 &         0.874  &        0.922    \\
      &  \textit{0.1219} & \textit{0.205} &\textit{0.193} &          & \textit{0.172} & \textit{0.081} &\textit{0.049}   \\
  \hline
\end{tabular}
\caption{Average $\text{cor}^2(\mathbf{X}\bm{\beta},\mathbf{X}\widehat{\mathbf{b}})$ across 10,000 simulated data sets for Model \ref{large_y_model} with different choices of $n$ and $p$ and where
 $\widehat{\mathbf{b}}$ is the estimate from various methods.  RR, RM1 and RM2 refer to Methods OLS, M1 and M2 but where robust regression $M$-estimation has been used in the regression step with the Huber weight function.  M1-trim and M2-trim
 refer to Methods M1 and M2 but where 10\% of observations with the largest Cook's distance were trimmed.  Standard deviations are in italics.}\label{table:sim_large_y}
\end{table}

In Table \ref{table:sim_large_y} we report the average squared correlations between the true and estimated dimension reduced predictors for 10,000 simulated runs from Model \ref{large_y_model} with standard deviations in italics.  For PHD, the very large response values often generated can result in extremely poor results.  However, for OLS \cite{GA&PR13} showed that using the rank of the response instead of the response itself could provide improved results.  Consequently, we used the rank of the response values for PHD and this approach provides vast improvements.  Therefore, the PHD results presented in this table are based on this estimation.  As well as employing OLS, PHD based on ranks and Methods 1 and 2, we also consider other variations that can be used to limit the influence of very large response values.  RR, RM1 and RM2 refer to the usual OLS, Methods 1 and 2 but where OLS has been replaced with the $M$-estimation robust version \citep{HU64, HU73} with the Huber weight function.  To do this we used the \texttt{rlm} function from the \texttt{MASS} package \citep{MASS} in R \citep{R}.  M1-trim and M2-trim refer to Methods 1 and 2 where 10\% of observations with the largest Cook's distance have been trimmed prior to the least squares step \citep[this is one of the trimming procedures from][]{PR08}.  The iterative estimation scheme for this model and methods did not provide improved results so for simplicity they have not been considered here.  Not surprisingly, OLS and the $M$-estimation equivalent completely fail even for large $n$ due to symmetric dependency.  Methods 1 and 2 perform much better but can still struggle as evident by the moderate average squared correlations and large standard deviations.  On the other hand PHD performs well, in particular for the larger sample size settings.  For Methods 1 and 2 coupled with $M$-estimation, we see improved performance over PHD for both methods.  These results suggest that by using the good PHD results in the transformation step to remove the symmetric dependency problem and then $M$-estimation to protect against large response values, excellent results can be achieved.  For the trimming approaches, improvements have been found when compared to standard Methods 1 and 2, however the results are a little worse than PHD and much worse than the transformation plus $M$-estimation methods.

\section{Inverse regression methods and multiple direction OLS}

\subsection{Inverse regression approaches}

The transformations discussed in Section \ref{section:methods} are certainly not limited to OLS and PHD.  Here we briefly discuss the use of Sliced Inverse Regression \citep[SIR,][]{LI91} and Sliced Average Variance Estimates \citep[SAVE,][]{CO&WE91}.  For brevity, we only briefly discuss these methods here and the reader is directed to the aforementioned articles for more detail.  Let $S_1,\ldots, S_H$ denote $H$ non-overlapping yet collectively exhaustive intervals covering the range of $Y$.  Let $\bm{\mu}_h = E(\mathbf{x}|Y\in S_h)$ $(h=1,\ldots,H)$ denote slice means and consider the matrix $\mathbf{V}=\bm{\Sigma}^{-1/2}\sum^H_{h=1}p_h(\bm{\mu}_h-\bm{\mu})(\bm{\mu}_h-\bm{\mu})^\top\bm{\Sigma}^{-1/2}$ where $p_h$ is the probability of $Y \in S_h$.  For $\bm{\gamma}$ denoting an eigenvector of $\mathbf{V}$ corresponding to a nonzero eigenvalue, \cite{LI91} showed that $\bm{\Sigma}^{-1/2}\bm{\gamma}$ is an element of the span of $\bm{\beta}_k$'s from the model in \eqref{Kmodel} provided  a $K$-direction version of Condition \ref{LDC} holds.  Consequently, if $\mathbf{V}$ is rank $K$ then SIR can recover a complete basis for the dimension reduction directions.  However, SIR suffers from the same problems with symmetric dependency as OLS \citep{LI91,CO&WE91} and is therefore a candidate for the same type of transformation.

\cite{CO&WE91} introduced SAVE which does not suffer from symmetric dependency issues.  It does, however, and additional to Condition \ref{LDC} require that $\text{Var}(\mathbf{x}|\bm{\beta}^\top_1\mathbf{x},\ldots,\bm{\beta}^\top_K\mathbf{x})$ is constant.  Both conditions are satisfied when $\mathbf{x}$ is normally distributed although both will often approximately hold in practice \citep{HA&LI93, LE13}.  Let $\bm{\Sigma}_h = \text{Var}(\mathbf{x}|Y\in S_h)$ $(h=1,\ldots,H)$ denote slice covariance matrices.  Then SAVE is carried out similar to SIR but where $\mathbf{M}=\sum^H_{h=1}p_h(\mathbf{I}_p-\bm{\Sigma}^{-1/2}\bm{\Sigma}_h\bm{\Sigma}^{-1/2})^2$ is used instead of $\mathbf{V}$. For some models SAVE requires large sample sizes to achieve good results.  However, we have found that a variation of SAVE that was proposed by \cite{ZH&ZH&FE10} called Cummulative Variance Estimation (CUVE) often provides excellent results.  For $\mathbf{z}=\bm{\Sigma}^{-1/2}(\mathbf{x}-\bm{\mu})$, CUVE estimates $E[\{P(Y \leq \widetilde{Y})\mathbf{I}_p - \text{Var}(\mathbf{z}\ I(Y\leq \widetilde{Y})\}^2]$ where $\widetilde{Y}$ is an independent copy of $Y$ and where $I(\cdot)$ is the indicator function taking the value 1 if its argument is true and zero otherwise.  Similarly, \cite{ZH&ZH&FE10} provided Cumulative Mean Estimation (CUME) which is variation of SIR and based on $E[E\{\mathbf{z}I(Y\leq \widetilde{Y})\}E\{\mathbf{z}I(Y\leq \widetilde{Y})\}^\top]$.  Neither method requires choosing $H$ and \cite{SH&PR11} showed that they can be successfully combined to obtain excellent results.  Consequently, we will also consider CUME following transformation based on the first CUVE direction.

For SIR and SAVE the user must specify the subranges of $Y$ used for `slicing' with the easiest approach to set $H$ equally probable slices.  In practice this equates to ordering the data by the magnitude of the response and allocating an (approximately) equal number of observations to each slice.  For estimation we use Rs \texttt{dr} package \citep{WE02} which, by default, uses $\max(8, p+3)$ for $H$.

\begin{table}[h!t]
\centering
\begin{tabular}{llllll}
  \hline
    Method & $n=100$ & $n=200$ & Method & $n=100$ & $n=200$ \\ \hline
  SIR & 0.066 & 0.065 & SAVE SIR (M2) & 0.247 & 0.835 \\
      & \textit{0.106} & \textit{0.107} &  & \textit{0.292} & \textit{0.266} \\
  CUME & 0.070 & 0.068 & SAVE SIR (M2-it) &  0.363 & 0.934 \\
   & \textit{0.090} & \textit{0.090} &  & \textit{0.401} & \textit{0.234} \\
  SAVE & 0.174 & 0.585 & CUVE CUME (M2) & 0.972 & 0.993 \\
   & \textit{0.190} & \textit{0.237} &  & \textit{0.034} &  \textit{0.003} \\
  CUVE & 0.889 & 0.957 & CUVE CUME (M2-it) & 0.987 & 0.995\\
  & \textit{0.053} & \textit{0.016} &  & \textit{0.021} & \textit{0.002} \\
  \hline
\end{tabular}
\caption{Average $\text{cor}^2(\mathbf{X}\bm{\beta},\mathbf{X}\widehat{\mathbf{b}})$ across 10,000 simulated data sets for Model \ref{large_y_model} with two different choices of $n$ and $p=20$ and where
 $\widehat{\mathbf{b}}$ is the estimate from various methods.  M2 refers to transformation Method 2 and M2-it refers to this transformation with iterative estimation.  Standard deviations are in italics.}\label{table:sim_large_y_SIR}
\end{table}

An advantage of SIR, SAVE, CUME and CUVE is that the $y_i$'s are used only to allocate $\mathbf{x}_i$'s.  Consequently, we would not expect extremely large $y_i$'s to have the same detrimental effect on estimation as they do for OLS.  To highlight this we reconsider Model \ref{large_y_model} and adopt Method 2 as follows.  SAVE is used to get the initial estimate to the direction of $\bm{\beta}$.  SIR is then used on either the transformed $y_i$'s (Method 1) or $\mathbf{x}_i$'s (Method 2) where the SAVE direction has been used to facilitate the transformations.  Similarly, we use CUME following a transformation using the CUVE estimated direction.  We also consider the iterative estimation procedures for both.  In Table \ref{table:sim_large_y_SIR} we provide the results from 10,000 simulations where $p=20$ and $n=100$ or 200.  Due to symmetry, both SIR and CUME fail to estimate the direction of $\bm{\beta}$.  SAVE also has trouble estimating the direction of $\bm{\beta}$, especially for $n=100$, although improvements are found for $n=200$ and we observed good results for $n=500$ (not shown).  CUVE, on the other hand, performs well, even for $n=100$.  The results also indicate that transformation Method 2 results in improved estimation although the combination of SAVE and SIR is only successful for the larger sample size.  The combination of CUVE and CUME, however, provides excellent results even for $n=100$.  For both approaches, the iterative estimation scheme also provides improvements, as evidenced by the increase mean squared correlations.

\subsection{Detecting multiple directions with OLS}

\cite{GA&PR13a} showed that OLS can be used to find multiple directions when different response transformations are employed.  They obtain several OLS slopes with with weights related to a leave-one-out sensitivity and then obtain one or more directional estimates for the $\bm{\beta}_k$'s.  Similarly, we show here that a simple two-step estimation procedure for OLS can work exceptionally well when OLS is faced with the task of finding two directions, one of which is expected to be non-detectable due to symmetric dependency.  As a matter of comparison we will also consider PHD$|$OLS which is an iterative version of PHD and OLS considered by \cite{SH&PR11}.  Here, the first direction estimated is the OLS slope.  Then PHD is used conditional on this OLS slope estimate already been detected so that only new information is found in the second direction.  The model we will focus on is given below which was also considered by \cite{SH&PR11}.

\begin{model} $Y = \sin(0.5\bm{\beta}_1^\top\mathbf{x}) + \cos(0.5\bm{\beta}_2^\top\mathbf{x}) + 0.3\varepsilon$\label{cos_sine_model} where $\bm{\beta}_1=[1, 2, -3, 0, \ldots,0]^\top$ and $\bm{\beta}_2=[1,1,0,-2,0,\ldots,0]^\top$.\end{model}

For the model above we will consider the performance of SIR, PHD, PHD$|$OLS and three new approaches based on Methods 1 and 2.  For these new approaches we will use the OLS slope vector as the first estimated direction and then use the transformation methods to estimate a second direction.

\begin{table}[h!t]
\centering
\begin{tabular}{lllllllll}
  \hline
  Method    &  $n=100$ &  & $n=200$ &  & $n=500$ & & $n=1000$ & \\ \hline
     & $\overline{r}_1$ & $\overline{r}_2$ & $\overline{r}_1$ & $\overline{r}_2$ & $\overline{r}_1$ & $\overline{r}_2$ & $\overline{r}_1$ & $\overline{r}_2$ \\ \hline
  SIR     &          0.776  &         0.277  &        0.872  &        0.326    &         0.952 &         0.448 &         0.977 &         0.615    \\
          &  \textit{0.142} & \textit{0.199} &\textit{0.088} &\textit{0.230}   & \textit{0.028}& \textit{0.269}& \textit{0.013}& \textit{0.264}   \\
  PHD     &          0.928  &         0.410  &        0.967  &        0.422    &         0.987 &         0.428 &         0.994 &         0.431    \\
          &  \textit{0.046} & \textit{0.231} &\textit{0.020} &\textit{0.235}   & \textit{0.007}& \textit{0.237}& \textit{0.003}& \textit{0.236}   \\
  PHD$|$OLS &          0.929  &         0.678  &        0.967  &        0.832    &         0.988 &         0.931 &         0.994 &         0.965    \\
          &  \textit{0.048} & \textit{0.207} &\textit{0.020} &\textit{0.109}   & \textit{0.007}& \textit{0.038}& \textit{0.003}& \textit{0.018}   \\
  OLS,M1  &          0.942  &         0.716  &        0.976  &        0.850    &         0.991 &         0.934 &         0.996 &         0.966    \\
          &  \textit{0.045} & \textit{0.172} &\textit{0.017} &\textit{0.086}   & \textit{0.006}& \textit{0.037}& \textit{0.003}& \textit{0.018}   \\
  OLS,M2  &          0.949  &         0.722  &        0.980  &        0.852    &         0.993 &         0.935 &         0.996 &         0.967    \\
          &  \textit{0.038} & \textit{0.171} &\textit{0.013} &\textit{0.085}   & \textit{0.004}& \textit{0.036}& \textit{0.002}& \textit{0.017}   \\
  \hline
\end{tabular}
\caption{Average first and second canonical correlations $(\overline{r}_1,\overline{r}_2)$ between $\mathbf{X}[\bm{\beta}_1,\bm{\beta}_2]$ and $\mathbf{X}\widehat{\mathbf{B}}$ across 10,000 simulated runs for data generated from Model \ref{cos_sine_model} with different choices of $n$ and $p$ and where
 $\widehat{\mathbf{b}}$ is the estimate from one of five methods; OLS, PHD, Method 1 (M1) and Method 2 (M2).  Standard deviations are in parentheses.}\label{table:sin_cos}
\end{table}

In Table \ref{table:sin_cos} we provide the simulated average first and second canonical correlations between $\mathbf{X}[\bm{\beta}_1,\bm{\beta}_2]$ and $\mathbf{X}\widehat{\mathbf{B}}$ where $\widehat{\mathbf{B}}$ is a $p\times 2$ matrix consisting of the first and second estimated directions.  A large average first canonical correlation, $\overline{r}_1$, indicates that the approach successfully detects the first direction.  Similarly, a large $\overline{r}_2$ is indicative of good performance in detecting the second direction.  SIR and PHD each are capable of finding one of the directions - for SIR the direction it is expected to find $\bm{\beta}_1$ and for PHD it is $\bm{\beta}_2$.  However, these methods do not perform well at finding the other direction. PHD$|$OLS is expected to find both and the average canonical correlations indicate this, although the method may have some trouble in estimating the second direction for $n=100$.  The transformation methods provide improvements in estimating both directions; certainly with respected to SIR and PHD and marginally better results than even PHD$|$OLS.

We could similarly use robust $M$-estimation regression methods here too.  Rather than repeat the simulation for similar results, we will choose this approach for the example considered in the next section.

\section{The Ozone data example}

\cite{LI92} considered the Ozone data from \cite{BR&FR85} which consists of 330 observations and eight predictors (e.g. wind speed, humidity etc., refer to Table 4 of Li 1992 for full list of predictors).  The response is atmospheric ozone concentration.  \cite{LI92} notes that the method Sliced Inverse Regression SIR finds a quadratic relationship (although not one that includes symmetric dependency) between the response and eight predictors and that almost an identical relationship can be found using least squares.  Using PHD, another direction is found that eluded SIR and which provides an ESSP that exhibits symmetric dependency.

In this example we will also consider the Ozone data example.  Let $Y$ denote the response variable and $\mathbf{x}$ denote the eight-dimensional vector of predictor variables.  As previously we let the sample data be denoted $\{y_i,\mathbf{x}_i\}^{330}_{i=1}$.  We will base our model on $\sqrt{Y}$ which, as we will see shortly, allows methods such as OLS, $M$-estimator regression methods and SIR to detect a linear relationship between the the response and predictors.  We will also use robust $M$-estimator as a robust least-squares method with the Huber weight function in the analysis.  For convenience we will refer to this method as RR.

\begin{figure}[h!t]
\centering
\includegraphics[scale = 0.9]{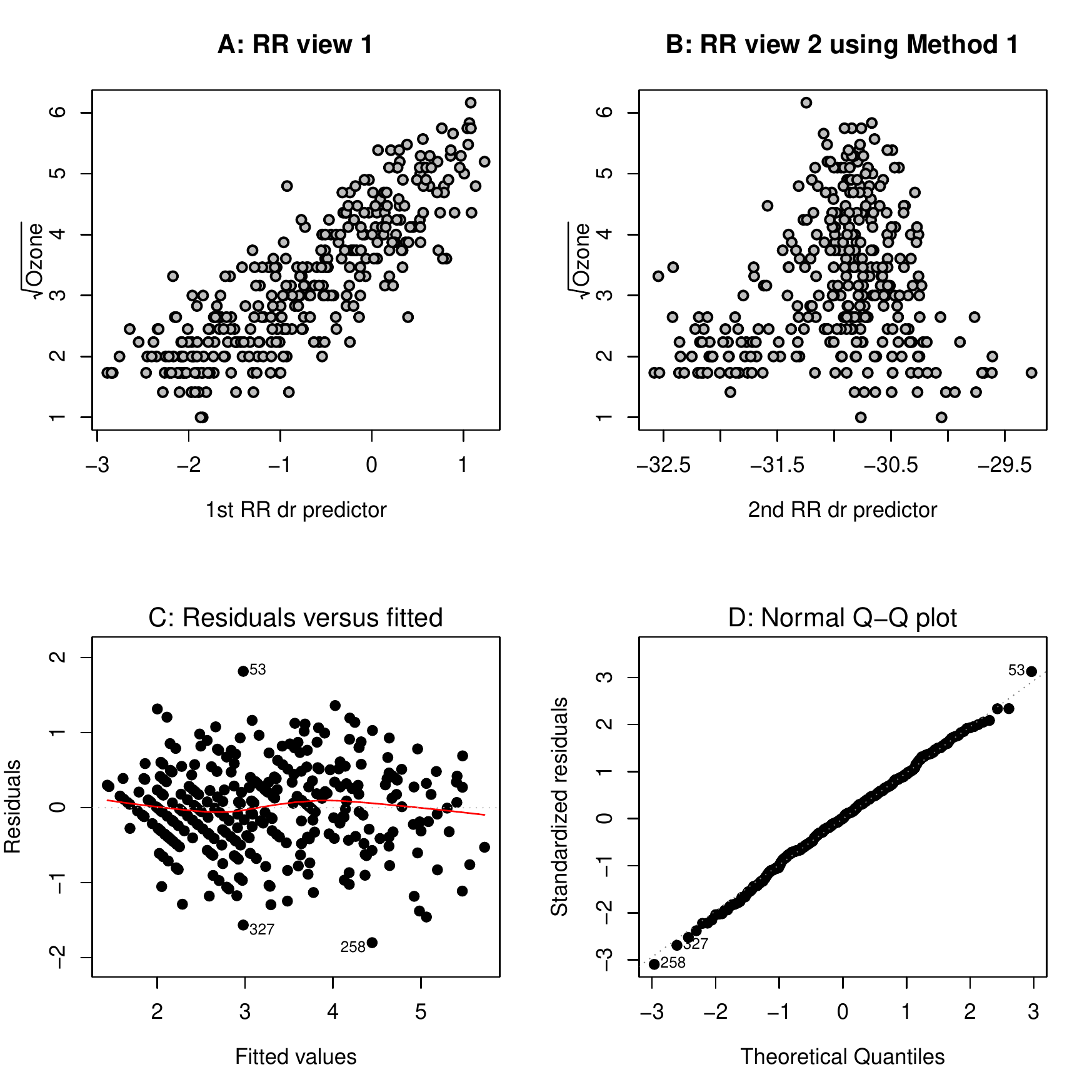}\caption{Plots of (A) the ESSP found by RR, (B) an ESSP created using a second direction found using RR following transformation Method 1, (C) residuals versus fits of a least squares fit to the dimension reduced predictors from Plots (A) and (B) and the square of these dimension reduced predictors and (D) the corresponding normal Quantile-Quantile plot.}\label{figure:3}
\end{figure}

Let $\widehat{\mathbf{b}}_1$ be the estimated slope for the RR regression of the $\sqrt{y_i}$'s on the $\mathbf{x}_i$'s.  A plot of the $\sqrt{y_i}$'s versus the $\widehat{\mathbf{b}}_1^\top\mathbf{x}_i$'s in Plot A of Figure \ref{figure:3} shows a linear relationship between the response and the dimension reduced predictors (labelled `1st RR dr predictor' on the plot).  We now use transformation Method 1 with the first PHD direction but with RR replacing OLS and let $\widehat{\mathbf{b}}_2$ denote this new estimate.  Plot B shows that RR has now found another direction exhibiting symmetric dependency.  We now use OLS to fit a model to the $\widehat{\mathbf{b}}_1^\top\mathbf{x}_i$'s, the $\widehat{\mathbf{b}}_2^\top\mathbf{x}_i$'s and the square of each of these (we did not included the multiple of the two for a full quadratic model since this had little contribution).  The estimated model is
$$\widehat{Y^{1/2}} = -212.89 + 1.22\times(\widehat{\mathbf{b}}_1^\top\mathbf{x}) + 13.94 \times (\widehat{\mathbf{b}}_2^\top\mathbf{x}) + 0.18\times (\widehat{\mathbf{b}}_1^\top\mathbf{x})^2 - 0.22\times (\widehat{\mathbf{b}}_2^\top\mathbf{x})^2.$$

The above fitted model explains approximately 75\% of the variation in square-root of the response indicating a good fit and all of the terms in the model were highly significant.  In Plots C and D we provide the residuals versus fits plot for the fit and also the Quantile-Quantile plot to check to see whether one could assume something close to a normal error term for the underlying model.  These plots are excellent indicating that if we were to assume a normally distributed error term with homogeneous variance then there is no evidence here to suggest that such an assumption would not hold approximately.  Consequently, we have successfully used RR twice to find two directions that can be used to construct a simple model with simple error term properties.

\section{Discussion}

This paper showed that simple response and predictor transformations can be used to remove the problem of symmetric dependency that effects some dimension reduction methods.  While we initially showed that OLS and PHD can be successfully employed in tandem for improved estimates, our approaches need not be limited to these methods.  To highlight this we also showed the popular robust $M$-estimation methods can be used as well as Sliced Inverse Regression in conjunction with Sliced Average Variance Estimates and associated cummulative slicing approaches.  These approaches are particularly useful when faced with very large response values that can be detrimental to OLS estimation.  Another interesting outcome from this paper was the ability in which OLS, and robust equivalents, could be used to find more than one direction.

\appendix

\section{Proof of Theorem \ref{theorem:tx(v)proof}}

Throughout let $E(\mathbf{x})=\bm{\mu}$ and $\text{Var}(\mathbf{x})=\bm{\Sigma}$ and recall $\mathbf{v}=c\bm{\beta}$. It can be shown that \cite[see, for e.g.,][]{PR05} Condition \ref{LDC} is equivalent to
\begin{equation}
E(\mathbf{x}|\mathbf{v}^\top\mathbf{x})=\bm{\mu}+\left\{(\mathbf{v}^\top\bm{\Sigma}\mathbf{v})^{-1}\mathbf{v}^\top[E(\mathbf{x}|\mathbf{v}^\top\mathbf{x})-\bm{\mu}]\right\}\bm{\Sigma}\mathbf{v}\label{LDC_eq}
\end{equation}
Since
$E[\text{sign}(\mathbf{v}^\top\mathbf{x} - \mathbf{v}^\top\bm{\mu})(\mathbf{x}-\bm{\mu})]= E\{\text{sign}(\mathbf{v}^\top\mathbf{x} - \mathbf{v}^\top\bm{\mu})E(\mathbf{x}-\bm{\mu}|\mathbf{v}^\top\mathbf{x})\}$, then from \eqref{LDC_eq},
\begin{equation}
E[t_x(\mathbf{x}-\bm{\mu}; \mathbf{v})]=E[\text{sign}(\mathbf{v}^\top\mathbf{x} - \mathbf{v}^\top\bm{\mu})(\mathbf{x}-\bm{\mu})]=\left\{(\mathbf{v}^\top\bm{\Sigma}\mathbf{v})^{-1}\mathbf{v}^\top E[t_x(\mathbf{x}-\bm{\mu}; \mathbf{v})]\right\}\bm{\Sigma}\mathbf{v}=c_4\bm{\Sigma}\mathbf{v}\label{Etx}
\end{equation}
where we identify here that $c_4\in \mathbb{R}$.

Similarly, by noting $E[\text{sign}(\mathbf{v}^\top\mathbf{x} - \mathbf{v}^\top\bm{\mu})(\mathbf{x}-\bm{\mu})Y]=E\left\{\text{sign}(\mathbf{v}^\top\mathbf{x} - \mathbf{v}^\top\bm{\mu})Y E(\mathbf{x}-\bm{\mu}|\mathbf{v}^\top\mathbf{x})\right\}$ since, from the model in \eqref{model}, $Y$ is a function of $\bm{\beta}^\top\mathbf{x}$ and $\varepsilon$ where $\varepsilon$ is independent of $\mathbf{x}$, we can also show that
\begin{equation}
\text{Cov}[\text{sign}(\mathbf{v}^\top\mathbf{x} - \mathbf{v}^\top\bm{\mu})(\mathbf{x}-\bm{\mu}), Y]=c_5\bm{\Sigma}\mathbf{v}\label{covtxY}
\end{equation}
for a $c_5\in \mathbb{R}$.  This shows that \eqref{ols-1} holds.

Now, using \eqref{Etx},
$$\text{Var}[t_x(\mathbf{x}-\bm{\mu}; \mathbf{v})]=\bm{\Sigma}-c_4^2\bm{\Sigma}\mathbf{v}\mathbf{v}^\top\bm{\Sigma}$$
since $E[t_x(\mathbf{x}-\bm{\mu}; \mathbf{v})t_x(\mathbf{x}-\bm{\mu}; \mathbf{v})^\top] = E[(\mathbf{x}-\bm{\mu})(\mathbf{x}-\bm{\mu})^\top] = \bm{\Sigma}$.  Therefore \citep[for e.g., use the Small Rank Adjustment Lemma,][page 19]{HO&JO85}
\begin{equation*}
\text{Var}[t_x(\mathbf{x}-\bm{\mu}; \mathbf{v})] =\bm{\Sigma}^{-1}+\frac{c_4^2}{1-c_4^2\mathbf{v}^\top\bm{\Sigma}\mathbf{v}}\mathbf{v}\mathbf{v}^\top
\end{equation*}
In conjunction with \eqref{covtxY}, this shows that \eqref{ols-2} also holds completing the proof.

\bibliographystyle{authordate4}
\bibliography{refs}

\end{document}